\documentclass[hyphens,sigconf,edbt]{acmart-edbt2022}

\def\BibTeX{{\rm B\kern-.05em{\sc i\kern-.025em b}\kern-.08em
    T\kern-.1667em\lower.7ex\hbox{E}\kern-.125emX}}

\usepackage{booktabs} 
\usepackage{verbatim}
\usepackage{listings}

\setcopyright{rightsretained}

\acmDOI{}

\acmISBN{978-3-89318-086-8}

\acmConference[EDBT 2022]{25th International Conference on Extending Database Technology (EDBT)}{29th March-1st April, 2022}{Edinburgh, UK} 

\acmYear{2022}

\renewcommand\footnotetextcopyrightpermission[1]{}

\begin{document}
\title{Desbordante: from benchmarking suite to high-performance science-intensive data profiler (preprint)}


\author{
George Chernishev\textsuperscript{1,2}, 
Michael Polyntsov\textsuperscript{2}, 
Anton Chizhov\textsuperscript{2}, 
Kirill Stupakov\textsuperscript{2},
Ilya Shchuckin\textsuperscript{2},
Alexander Smirnov\textsuperscript{2}, 
Maxim Strutovsky\textsuperscript{1,2}, 
Alexey Shlyonskikh\textsuperscript{2}, 
Mikhail Firsov\textsuperscript{2}, 
Stepan Manannikov\textsuperscript{2},
Nikita Bobrov\textsuperscript{2},
Daniil Goncharov\textsuperscript{2},
Ilia Barutkin\textsuperscript{2},
Vladislav Shalnev\textsuperscript{2},
Kirill Muraviev\textsuperscript{2},
Anna Rakhmukova\textsuperscript{2},
Dmitriy Shcheka\textsuperscript{2},
Anton Chernikov\textsuperscript{2},
Dmitrii Mandelshtam\textsuperscript{2},
Mikhail Vyrodov\textsuperscript{2},
Arthur Saliou\textsuperscript{2},
Eduard Gaisin\textsuperscript{2},
Kirill Smirnov\textsuperscript{2}}
\affiliation{%
  \institution{\textsuperscript{1}Unidata LLC, Russia \\ \textsuperscript{2}Saint-Petersburg University, Russia}
  \city{Saint-Petersburg, Russia} 
}
\email{
georgii.chernyshev@unidata-platform.ru,
polyntsov.m@gmail.com,
anton.i.chizhov@gmail.com,
kirill.v.stupakov@gmail.com,
shchuckinilya@gmail.com,
alexander.a.smirnovv@gmail.com,
maxim.strutovskiy@unidata-platform.ru,
shlyonskikh.alexey@gmail.com,
mikhail.a.firsov@gmail.com,
manannikov.st@gmail.com,
nikita.v.bobrov@gmail.com,
goncharov.y.daniil@gmail.com,
ilia.d.barutkin@gmail.com,
vladislav.a.shalnevv@gmail.com,
rakhmukova.anna@gmail.com,
dmitriy.v.shcheka@gmail.com,
chernikov.a.anton@gmail.com,
mandelshtamd@yandex.ru,
mikhail.v.vyrodov@gmail.com,
arthur.salio@gmail.com,
edu.gaisin@gmail.com,
kirill.k.smirnov@gmail.com
}

\renewcommand{\shortauthors}{}

\begin{abstract}

Pioneering data profiling systems such as Metanome and OpenClean brought public attention to science-intensive data profiling. This type of profiling aims to extract  complex patterns (primitives) such as functional dependencies, data constraints, association rules, and others. However, these tools are research prototypes rather than production-ready systems. 

The following work presents Desbordante~--- a high-perfor\-mance science-intensive data profiler with open source code. Unlike similar systems, it is built with emphasis on industrial application in a multi-user environment. It is efficient, resilient to crashes, and scalable. Its efficiency is ensured by implementing discovery algorithms in C++, resilience is achieved by extensive use of containerization, and scalability is based on replication of containers. 

Desbordante aims to open industrial-grade primitive discovery to a broader public, focusing on domain experts who are not IT professionals. Aside from the discovery of various primitives, Desbordante offers primitive validation, which not only reports whether a given instance of primitive holds or not, but also points out what prevents it from holding via the use of special screens. Next, Desbordante supports pipelines~--- ready-to-use functionality implemented using the discovered primitives, for example, typo detection. We provide built-in pipelines, and the users can construct their own via provided Python bindings. Unlike other profilers, Desbordante works not only with tabular data, but with graph and transactional data as well.

In this paper, we present Desbordante, the vision behind it and its use-cases. To provide a more in-depth perspective, we discuss its current state, architecture, and design decisions it is built on. Additionally, we outline our future plans.

\end{abstract}

\graphicspath{ {./images/} }

%
%



\maketitle

\section{Introduction}

According to~\cite{10.1007/s00778-015-0389-y}, data profiling is the ``set of activities and processes to determine the metadata about a given dataset''. Such metadata can be useful for dataset exploration, various tasks related to data quality, database management and database reverse engineering. It also has many applications~\cite{10.5555/3312004, 10.1145/3310205} in data integration and query optimization domains.

Data profiling can be divided into naive and science-intensive. Naive profiling concerns extraction of such simple facts as number of rows and columns in a table, minimum and maximum values in a column, detection of atomic column data types and so on. There are hundreds of data profiling tools that belong to this class, as almost all big information system vendors offer them. Many open-source tools exist as well, and one of the most prominent ones is Pandas Profiling~\cite{pandasProfiling}. 

On the other hand, science-intensive profilers focus on extraction of complex metadata, which requires sophisticated algorithms. Examples of such metadata discovery are the following: extraction of all kinds of functional dependencies from tables, both exact and relaxed~\cite{7219433}, association rule mining~\cite{10.5555/2677098}, detection of semantic column types~\cite{10.1145/3292500.3330993}, discovery of data constraints~\cite{denialConstraints, algebraicConstraints} and many more. Tools that offer such profiling are much rarer. Two significant systems that offer such functionality are Metanome~\cite{10.14778/2824032.2824086} and OpenClean~\cite{10.14778/3476311.3476339}.

In this paper, we present Desbordante (Spanish for \textit{boundless})~--- a high-performance science-intensive data profiler. It is inspired by Metanome, but differs from it in various key points. First of all, the focus of Desbordante is on industrial applications in a multi-user environment. Extracting complex metadata requires sophisticated algorithms which are very computationally expensive and crash-prone. Desbordante addresses these issues by taking a mindful approach to algorithm implementation and via specifically-designed application architecture.

Second, we have a different vision of use-cases for such a tool. We envision our primary users as domain experts who possess a large amount of data that they would like to explore, and at the same time, they are not necessarily IT professionals. Our users wish to discover various patterns in their data which state a non-trivial fact. These patterns are formally described using a variety of the so-called primitives. By itself, a primitive is a description of a rule that holds over the data (or a part of it), described formally, by mathematical methods. Functional dependencies can be considered as a well-known example.

The kinds of experts who could be interested in pattern discovery are:
\begin{enumerate}
    \item Bioinformatics researchers, chemists, geologists, and in fact almost any scientist working with large amounts of data, especially those obtained experimentally.
    \item People working with financial data: financial analysts, salespeople, traders, who all also have a lot of data at hand that can be explored.
    \item Data scientists, data analysts, machine learning specialists.
\end{enumerate}

For scientists that work with large amounts of data, finding a primitive indicates the presence of some pattern. Based on it, they may be able to formulate a hypothesis or even draw conclusions immediately (if there is enough data). At the very least, the found pattern can give them a direction for further study. For example, the bioinformatics group of the Saint-Petersburg JetBrains lab used such primitives in their work~\cite{10.1145/3459930.3469499}.

In the case of financial data, the researcher can also try to obtain some kind of hypothesis (for example, ``out of all cars offered by a competing company, red ones are the best selling''). However, there are more mundane and more in-demand applications: cleaning errors in data, finding and removing inexact duplicates, and many more. Note that scientists might also be interested in this functionality, albeit to a much lesser extent.

As for machine learning, the found primitives can help in feature engineering and choosing the direction for the ablation study.

The aforementioned use-cases lead to the following specific requirements for our primitive discovery tool:
\begin{itemize}
     \item Focus on approximate primitives. Our users work with real data, therefore implementing approximate primitives should be of priority.
     \item Focus not only on primitive discovery, but also on primitive validation and explainability of results. Our users need to be provided with information why a particular instance of a primitive does not hold.
     \item Focus on tunability. Our users wish to fine-tune discovery process by specifying various constraints on sought-after primitives.
     \item Focus on non-tabular data. While most popular data type are tables, our users are also interested in other types such as graphs and transactional data.
     \item Focus on supporting multiple interfaces. Some of our users need console application, some a rich web UI, and some need a Python interface.
\end{itemize}

The academic database community has created a large amount of primitives describing many different patterns that may be present in data. For example,  there are more than thirty different formulations only for the class of relaxed functional dependencies~\cite{7219433}. And every year novel primitives continue to appear. However, these primitives are largely unknown to people outside of the community. In the worst case they simply remain a theoretical result, and at best they exist in the form of a little-known prototype which is not ready for industrial application. There are several tools~\cite{10.14778/2824032.2824086, 10.14778/3476311.3476339} that contain collections of algorithms that discover and validate primitives but they are not production-ready, too. They are  performance-bound and fail to cater to specific needs of our users by lacking required functionality.

Therefore, the idea of Desbordante is to ``open'' these primitives to the general public and give everyone the opportunity to study their data.

Now, turning to the Desbordante itself:

\begin{itemize}
    \item Industrial focus is expressed through efficiency and resilience. Unlike other science-intensive profilers the kernel of Desbordante is implemented in C++.
    \item Desbordante comes with a console and a web version. Also, Python interface is provided.
    \item It supports pipelines~--- ready-to-use functionality implemented using the discovered primitives, e.g, typo detection. Users can construct their own pipelines via python bindings.
    \item It supports tabular and non-tabular data (graph, transactional).
\end{itemize}

Desbordante is the open-source\footnote{ https://github.com/Mstrutov/Desbordante (algorithms), \\ https://github.com/vs9h/Desbordante (web application).} project implemented using the modern tech stack. You can try the deployed demo here\footnote{https://desbordante.unidata-platform.ru}.

The main contribution of this paper is the description of the tool vision, architecture, and approaches we took to satisfy user needs and ensure its high performance, resilience, and scalability. We also discuss tool positioning, list currently supported primitives, and present future plans.

The paper is organized as follows. In Section~\ref{sec:relwork} we present related work and discuss existing profiling tools. Next, in Section~\ref{sec:primitives} we describe primitives and requirements specific to their discovery. In Section~\ref{sec:desb} we examine Desbordante from a 10K feet view and below by describing project's vision, core functionality, considerations regarding user experience and system's performance. In Section~\ref{sec:architecture} we sketch a system architecture of Desbordante, technology stack, hows and whys of implemented microservices. In Section~\ref{sec:future} we list our future plans and milestones to achieve.

\section{Related Work}\label{sec:relwork}

It is hard to classify existing tools in such a way that they would exactly match with Desbordante by properties,  functionality, or vision. Therefore, we review most well-known tools which implement various data profiling techniques and can mine primitives or at least highly rely on them.

The closest relative to Desbordante in the data profilers family is Metanome~\cite{10.14778/2824032.2824086}. In a nutshell, Metanome is a framework which provides developers with an infrastructure and a corresponding set of interfaces for implementing and benchmarking primitives. Metanome's architecture makes the process of developing and testing research ideas as fast and convenient as it can be, thus enabling developers to concentrate on the algorithms instead of boilerplate code for database connectors, text processing utilities, etc. Metanome was developed by the Hasso Plattner Institute group and almost any algorithm developed by the same group can be plugged into Metanome as a JAR file (the concept of an algorithm as connectable compiled code is what Desbordante currently lacks). However, Metanome can not be considered as a true industrial alternative to Desbordante due to the reasons we present in an extensive evaluation and comparison of both platforms~\cite{9435469}: larger memory footprint, inferior dependency discovery speed, just to name a few. Still, Metanome system is an inspiration for Desbordante as for a user-friendly, high-performance and flexible data profiler.

The goal of the OpenClean~\cite{10.14778/3476311.3476339} system is to become a part of a modern data science stack by taking a niche of data cleaning and profiling. Being an open-source Python library, OpenClean provides its users with an environment where they can seamlessly integrate data profiling with other frameworks and libraries for data processing and machine learning.
The concept of FD is used in two ways: checking data for FD violations and FD mining. The former is implemented in Python as a combination of mapper and group filters (much like the \lstinline{SELECT}\ldots \lstinline{GROUP BY} \ldots \lstinline{HAVING} idiom for FD checking in SQL). For records which violate the FD, a repair process can be started via OpenClean repair strategies or user-defined ones. The primitive mining functionality is provided by a standalone package which initiates a subprocess for running Metanome JAR files. Basically, any algorithm that was once implemented for Metanome, can be run within OpenClean. It also means that in terms of algorithm performance OpenClean inherits all problems of Metanome.

Unlike OpenClean, most data cleaning tools have no built-in functionality for primitive mining, and expect a user to provide primitives as an input data to the cleaning process~\cite{HoloClean, Horizon, SAMP, Holistic}. For example, a data repairing framework HoloClean~\cite{HoloClean} makes good use of denial constraints (DC), which subsumes FD, CFD and matching FD concepts. The input of HoloClean is an inconsistent dataset, a set of DCs and any external knowledge which can be used to repair the dataset. HoloClean combines every piece of available information and proposes solutions that can bring the data to a consistent state. Since the main goal of HoloClean is to pave a road to a careful restoration of a consistent form of a dataset, the tool does not implement any internal mechanisms for mining of primitives or for efficient in-place inference of metadata. The same vision is shared by Horizon~\cite{Horizon}, which computes a FD pattern graph based on the FD set provided, and constructs a solution via pattern graph static analysis.

However, not every data profiling tool considers primitives as a great deal for a cleaning process or error detection task. Otherwise, they rely on machine learning, probabilistic methods~\cite{SCARE}, or a curated knowledge base like Katara does~\cite{KATARA}. The authors of Katara even refuse to consider FDs as trusted metadata, since this type of primitive can not guarantee that data would be fixed in a non-ambiguous way.

Some data engineering tools follow a different philosophy: instead of fixing data, they make sure it is tidy in the first place. Great Expectations~\cite{greatExpectations} allows its user to define complex integrity constraints which are used as an assertion mechanism while the data flows through ETL processes. Such tiny unit tests for data validation can be embedded into a workflow and immediately raise a flag if anything unexpected happens, e.g., newly arrived data violates a primitive that was described within the Great Expectations framework. A similar idea is used in the Auto-Validate~\cite{10.1145/3448016.3457250} system.

The aforementioned tools are implementations of research findings which are carefully surveyed in dozens of papers, and some of these tools are open-source and free to use. To make the overview complete, we would like to list some of industrial solutions. 

Most commercial solutions provide support for the concept of FD as a part of data profiling: SAP Information Steward, Oracle Warehouse Builder, Informatica Data Quality, Microsoft SQL Server Data Profiling Task, and Talend Open Studio can return functional dependencies which almost hold on data, or verify whether a user-specified dependency holds. For each AFD, these tools also maintain the fraction of records which violate a dependency. It seems that this way of processing FD/AFD is almost a must-have feature for any data profiling tool and it comes handy when performing anomaly detection and exploring ``broken'' records. However, usually FD/AFD are the only types of primitives that are implemented in a pay-to-use tool, since nowadays their focus has shifted to the machine learning side of data profiling spectrum.

\section{Discovery of science-intensive primitives}\label{sec:primitives}

\subsection{Current State and Motivation}

By itself, a primitive is some description of a rule (a pattern) that holds over the data, described mathematically. Functional dependencies can be a good example: dependency A $\longrightarrow$ B (A and B being columns) holds if for each pair of rows it is true that from the equality of values in A follows the equality of values in B.

There are several hundred of types of primitives~\cite{9302878, 7219433}, and each of them has well-established properties and a sound theory behind it. New types are developed all the time, too.

However, as stated in the Introduction, they largely stay within the database and associated communities, they provide no benefit to broader public.

The reasons for this are the following:
\begin{itemize}
     \item Largely, implementations of primitives are poorly accessible or not available at all:
     \begin{itemize}
         \item The majority of them was developed in the pre-Github era when it was not customary for authors to provide source code or the source code was published on research group's web site, which is usually long-dead now. Either way, currently there is no source code available.
         \item These that are accessible now are scattered around the Web, on personal web sites or in obscure repositories. Of course, the presenting paper usually includes a link, but prospective users have to know about the primitive and the paper first, which is not the case.
     \end{itemize}

     \item If they are available, they are hard to set up and run. For example, the newcomers of our team took from 6 to 12 work hours to set up and run Metanome. And at the same time they are mostly computer science students who familiar with IT specifics. Thus, for a non-IT specialist who would like to try some primitive, it will be a rather tedious task.

     \item Each available implementation of a primitive (or even discovery algorithm) would require its own software ecosystem to set up and maintain. It is an another obstacle to overcome for a prospective non-technical user who would like to try some primitives.
     
     \item Finally, available implementations are usually proofs of concept or prototypes which were made for some paper and were abandoned later. Therefore, they are usually not very efficient since they were developed in language which favours rapid prototyping, like Java or Python. These languages lack efficiency and low-level tunability of C++. There are also scalability issues in a sense that real world datasets are likely to be larger than those benchmarked by paper authors. Moreover, these implementations may crash when processing a dataset which was not benchmarked by its authors. Thus, it is necessary to shift the limits of applicability further (since the name~--- Desbordante).
\end{itemize}

\begin{figure*}[t]
\includegraphics[width=0.85\linewidth]{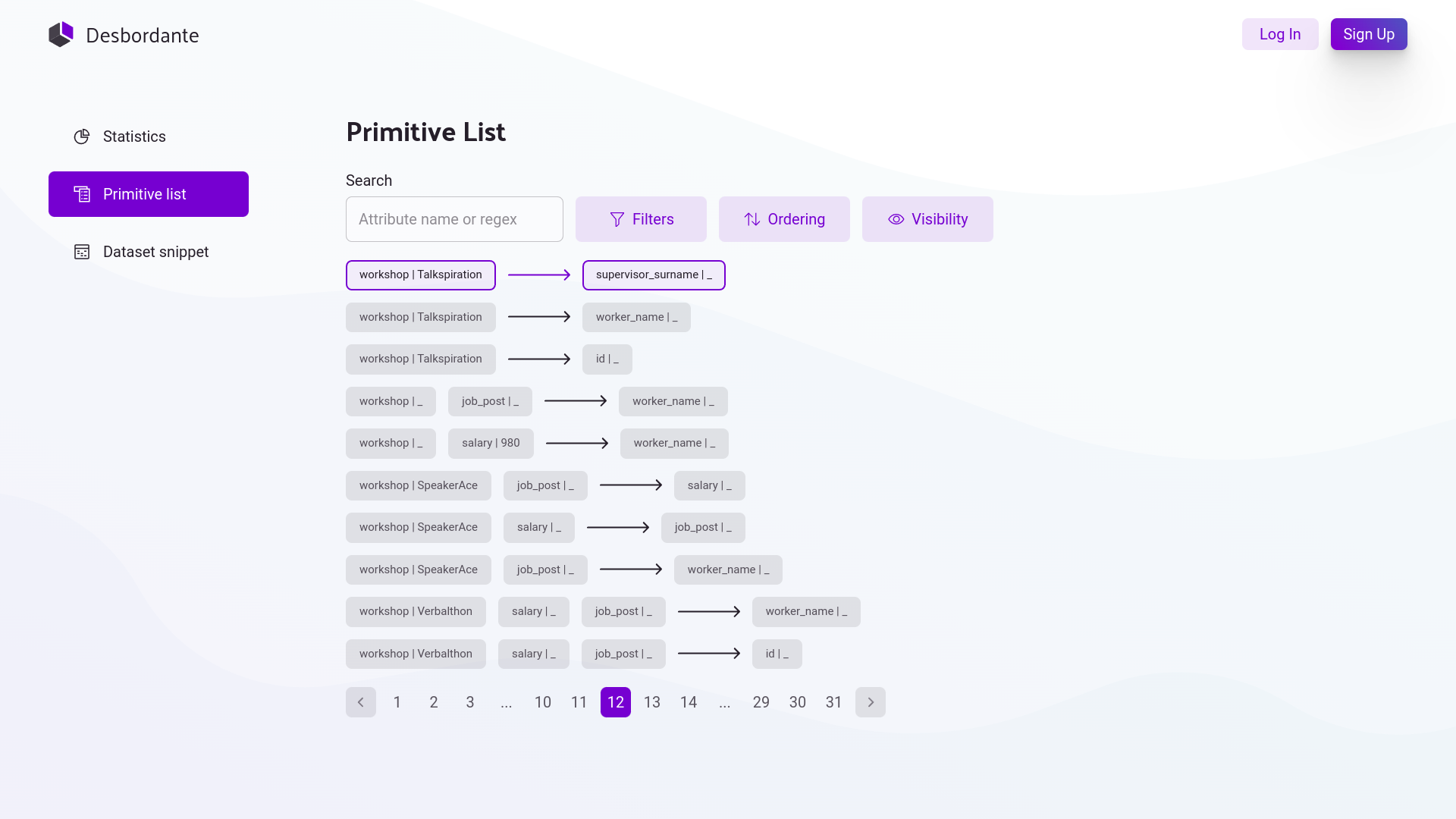}
\caption{The result page of the web version of Desbordante (conditional functional dependencies)}
\label{fig:demo}
\centering
\end{figure*}

There are platforms which try to address these issues, such as Metanome or OpenClean. However, they fail to address all these issues at once. Thus, there is a need for an industrial-grade platform which will open primitive discovery to the broader public and Desbordante tries to achieve this goal.

\subsection{Specifics of science-intensive primitive discovery}

Discovery of science-intensive primitives has its own specifics, which can be described by several aspects, divided into two groups. The ones belonging to first group are inherent to all science-intensive profilers and stem from the nature of primitive discovery task.

\begin{enumerate}
    \item Primitive discovery is a computationally hard problem. Discovery algorithms run into time or memory limits even for small datasets. Consider, for example, Table 1 from \cite{10.1145/2882903.2915203} where one can see sizes of datasets that can be mined for functional dependencies using server-class hardware. All datasets except two are smaller than one megabyte. The situation is similar in case of other primitives.  Therefore, in order to make the discovery of primitives truly usable, we need to address these limitations.
    
    \item Implementations of discovery algorithms are very complex, frequently depend on third-party libraries, and in general they solve a task belonging to the forefront of science. Therefore, they are fragile~--- they can crash or freeze on some inputs. Therefore, when ``industrializing'' them, one has to improve reliability of the application by making it fault-tolerant.
\end{enumerate}

The second group describes aspects which are specific to the vision and goals of our system. These reflect use-cases and needs of our users.

\begin{enumerate}

    \item Our users are more interested in approximate primitives. Real world data is likely to have all kinds of errors, missing values, and other types of artifacts. Therefore, exact versions of primitives are not applicable, they will be rarely found in real data. Instead, developing our profiler, we must provide inexact versions, which will allow some degree of error.

    \item Our users need not only discovery of primitives, but also their validation. Unlike discovery, validation accepts a specific instance of a primitive (e.g. a specific functional dependency) as input and returns whether it holds or not. This leads to the need for special screens in which the user can analyze the data and see what prevents a given primitive from holding (e.g. conflicting values, rows, etc.).
    
    \item Our users need to be provided with various tuning knobs that will govern the discovery process. For example, concerning the discovery of functional dependencies, it is well-known that dependencies with a larger left-hand side are less valuable. Their discovery usually does not indicate the presence of a real dependency, but instead points to the fact that the data segment which was used for mining is too small to contain a counterexample. Primitive discovery process is always costly and it is worthwhile to skip unnecessary computations. Another important example is setting the error threshold for approximate primitives. At the same time, correct values depend on the particular dataset and user goals.
    
    \item Our users have different preferences regarding the interface to use. Some of them prefer an old-school command-line interface, while others ask for a rich web UI. Furthermore, in order to open primitives to data scientists it is essential to provide a Python interface.
\end{enumerate}

Desbordante aims to take into account these specifics.

\begin{figure*}[t]
\includegraphics[width=\linewidth]{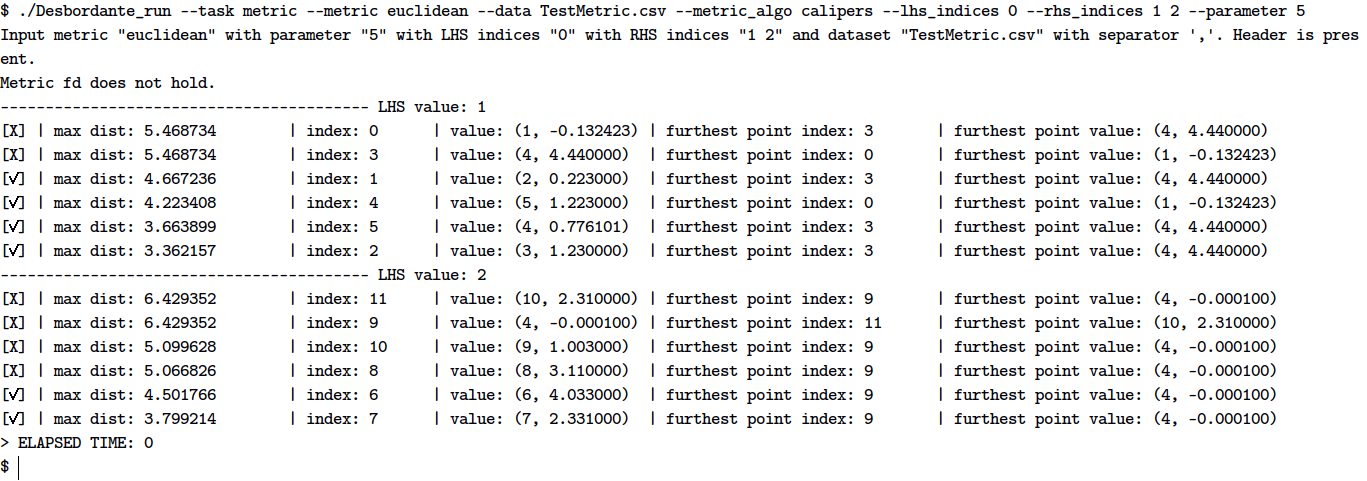}
\caption{Console validation screen for metric functional dependency}
\label{fig:screen}
\centering
\end{figure*}

\section{Desbordante}\label{sec:desb}

\subsection{Overview}

The core of Desbordante is a C++ library containing all auxiliary data structures needed for primitive discovery algorithms, the algorithms themselves and all required surrounding infrastructure. The library provides an API for executing the algorithms and obtaining their results which is used by the back-end of the web application. There is also additional library version with python bindings,
so that all Desbordante features can be used from within Python. 
There are two available user interfaces for the Desbordante library: console and web interface.

Desbordante supports two different types of tasks (and therefore algorithms):

\begin{itemize}
    \item Discovery: find all holding instances of specified primitive over specified dataset;
    \item Validation: given a primitive instance determine whether it holds over specified dataset, provide additional information about what prevents it from holding otherwise.
\end{itemize}

The usage workflow is generally the following:

\begin{enumerate}
    \item Select the primitive and desired algorithm for its discovery or validation;
    \item Specify a dataset to work on and the required parameters;
    \item Execute the algorithm and retrieve the results;
    \item Filter and sort the results as needed.
\end{enumerate}

These steps are clearly separated in the web application, while in the console version all the required information for the algorithm (steps 1--2) can be set directly via CLI parameters. Since the web application was designed to be used by non-IT professionals in the first place, it should provide quality of life features besides its main functionality. Examples of such features are a viewable snippet of the selected dataset, a user-friendly interface with extensive usage examples, ready-to-use pipelines, and a progress bar which shows the current execution status of a task. Desbordante itself and the web application have a set of the built-in datasets. They mainly serve two purposes. First, help users to understand what possible data insights they can get using specific primitives. Second, make newbies familiar with the workflow of the tool. In addition to built-in datasets, the web application allows the user to upload their own. These datasets constitute user's personal library which shares the same workflow as the built-in ones.

\subsection{User-facing aspects}

The first feature that we would like to present is the tunability of primitive discovery and validation processes. Each primitive discovery or validation algorithm has a set of various options. These options can be divided in two groups: general and primitive-specific parameters. General parameters are the properties which need to be set up for any discovery task: dataset, its delimiter, and the Boolean switch which indicates whether the dataset has a header. Then, there are primitive-specific parameters, the first and most important of which is the algorithm. For some primitives, the best performing algorithm is more or less known, but for some it is not. Moreover, a discovery algorithm may perform badly on a ``wide'' or a ``long'' table. It may also crash due to the specifics of a particular dataset since different algorithms are built upon different principles and, for example, may require too much memory. Therefore, we have decided to provide several available algorithms and in some cases all of them.

Each of these algorithms has its own set of supported parameters. First of all, they are what to validate or to look for, some filters on the primitive instance.
Then, if an algorithm supports discovery (or validation) of an approximate primitive, then the degree of allowed violations. Next, for discovery algorithms it is useful in some way to limit the ``depth'' of search. Discovery process is time-consuming and, at the same time, all instances are not always needed. Finally, a user can specify the number of threads which will be used for primitive discovery or validation if the selected algorithm supports multi-threading.

The next important feature are custom screens for the primitive validation task. This task reports whether a specific primitive holds or not. However, if it does not hold, users need explanations and answers to questions such as ``how much is lacking?'' and ``what prevents it from holding?''. It is essential to provide such information since it is an important knowledge about data being explored. It can indicate errors in data and point out ``problematic'' records. Therefore, there is a need for a screen that will provide this information. 

An example of such a screen for the console version of metric functional dependency~\cite{4812519} validation is presented in Figure~\ref{fig:screen}. It shows clusters of records that share the same left hand side, but differ in the right one. The ``x'' marks records which are too far from the rest in terms of their right hand side. Their distance to any of points from the same cluster is larger than the specified one and therefore they are good candidates to be outliers. The user can sort clusters and records within clusters using various parameters such as distance, index, number of outliers and so on.

Primitive discovery task also implements result screens with rich interaction tools that allow sorting using various parameters, filtering with regular expressions, and so on.

\subsection{Pipelines}

Aside from primitive discovery and primitive validation tasks, Desbordante offers pipelines. A pipeline is a set of ready-to-use functionality implemented using discovered primitives, which benefits a non-expert end user. While discovered primitive instances are useful by themselves, we believe that it is important to demonstrate what can be done using them. 

There are two types of pipelines in Desbordante: built-in and custom. The first ones are present in the web application and have a rich interface. As a demo, we have included a typo detection pipeline in the deployed version. Its idea is as follows: 

\begin{enumerate}
    \item Find functional dependencies which almost hold, i.e. approximate dependency~\cite{10.14778/3192965.3192968} holds, but not the exact one. Present them to the user for inspection.
    \item Then, for the dependency selected by the user, present its clusters~--- row groups with the same left hand side and different values in the right hand side. These are the sets of rows which prevent exact FD from holding. In this screen, user can inspect the differences in the right hand side and decide whether the there is a typo or not. Having resolved the conflict for a particular cluster, a user can reupload the new version of dataset and continue data cleaning.
    \item In order to reduce the number of presented clusters, the user interface contains several parameters that enable cluster filtering. A threshold for dependency to be considered as ``almost'' holding can also be set.
\end{enumerate}

Note that the CLI does not provide built-in pipelines, since they require extensive interactivity.

While built-in pipelines are useful as a demo, they require significant effort to implement. As we are limited in resources, we put only the most useful scenarios on the web version. At the same time, we would like to allow users to experiment and build their own pipelines. For this, we provide an ability to build custom pipelines.

Contemporary data scientists use Python, and therefore it is essential to enable calling primitive discovery and validation tasks from Python programs.  For this, we employed the pybind11~\cite{pybind11} library to provide the necessary operators and data structures. Using these bindings, our users can call Desbordante algorithms to experiment and construct their own pipelines. We plan to add popular ones to the web version and develop a user-friendly interface for them.

\begin{figure}[t]
\includegraphics[width=\linewidth]{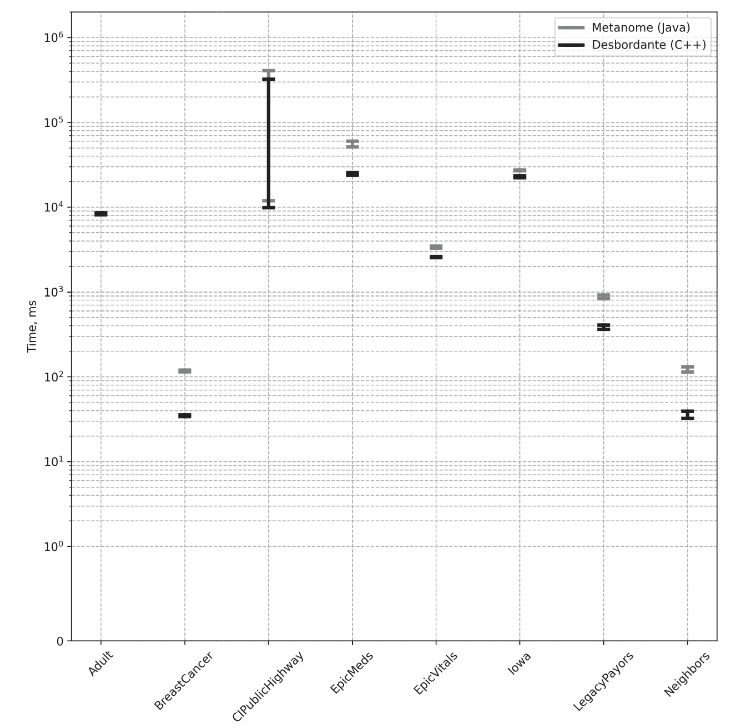}
\caption{Performance comparison of Desbordante and Metanome using Pyro}
\label{fig:comparison}
\centering
\end{figure}

\subsection{Performance}

Unlike all existing open-source solutions, the discovery part of Desbordante is fully implemented in modern C++. While popular languages such as Python and Java are relatively simpler and thus offer a fast development process, they possess a number of no less prominent drawbacks:
\begin{enumerate}
    
    \item Given equal effort put into code, the resulting performance of Java/Python applications is worse than that of C++, on average. 
    
    \item Java application performance can be unpredictable. Since explicit memory management is not possible in Java, programs rely on an automatic garbage collector, which may be invoked at any time. Therefore, run times may significantly differ even for consecutive invocations of single-threaded programs. 
    
    \item Java programs usually leave a higher memory footprint than C++.
    
    \item Finally, these languages restrict opportunities for low-level optimizations, such as vectorization via SIMD instructions. It is a critical drawback for solving a high-performance computing task.
\end{enumerate}

To demonstrate the validity of our arguments we have experimentally compared Desbordante with Metanome~\cite{9435469}. For this, we have selected the Pyro algorithm~\cite{10.14778/3192965.3192968} since it was one of the most promising primitives for the intended application scenarios. This algorithm discovers approximate functional dependencies.

The results are presented in Figure~\ref{fig:comparison}. The obtained improvement ranged from 1.19 to 3.43 times and was 2.12 on average. While the numbers are not really high, it is still an important result for such a computationally expensive problem.

Another significant benefit is the reduction of memory consumption~--- the memory footprint of Desbordante is approximately two times lower than Metanome's. This is crucial since many primitive discovery algorithms are memory-bound~\cite{10.1145/2882903.2915203}. Thus, reducing memory footprint enables the processing of larger datasets.

\begin{figure*}[t]
\includegraphics[width=\linewidth]{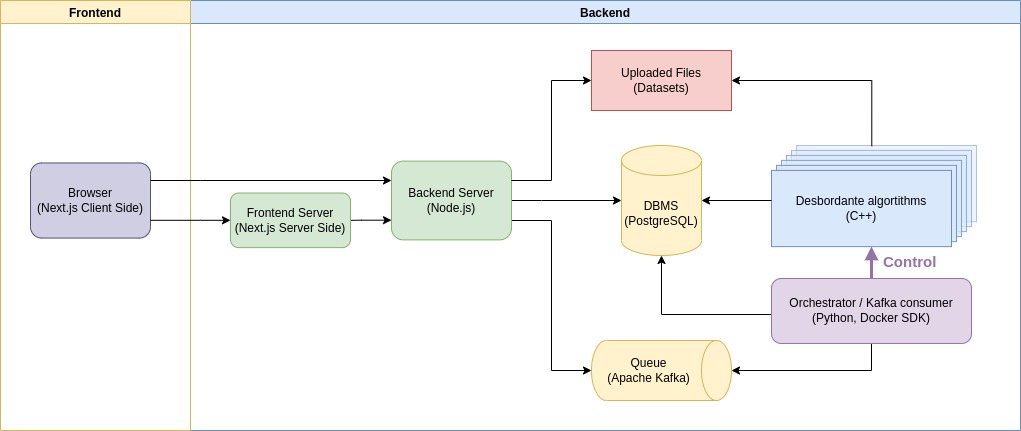}
\caption{Desbordante architecture in detail}
\label{fig:arch}
\centering
\end{figure*}

It is important to note that we have not exhausted the tuning potential of the C++ implementation. No sophisticated techniques were used (e.g., vectorization), no source code profiling was done, standard data structures and libraries were used, etc. Currently, Desbordante uses default C++ and Boost data structures, and we have not tuned their parameters. Desbordante does not rely on custom memory management libraries (allocators), but instead uses the C++ default. It is a well-known fact~\cite{MemoryAllocators} that using a special allocator is a simple yet efficient way to improve performance of C++ programs. Therefore, it is possible to improve performance even more.

Finally, we must also discuss the approaches that rely on distribution of  Java/Python code. Firstly, we believe that they will not improve the situation much. Primitive discovery problems are generally poorly scalable and naive approaches are not functional at all (e.g. see Fig. 1 in ~\cite{10.14778/3342263.3342638}). The reason for this is that it is necessary to pass over the whole (or a significant part of) dataset in order to get to the answer. Therefore, we believe that it is important to get the maximum performance out of single-node processing.

Secondly, since distribution approaches usually consist of a data shuffling scheme and some local algorithm, we can say that a distribution-based approach is not competing, but complementary to ours.

\subsection{Supported primitives}

Due to the reasons stated in Section~\ref{sec:primitives} Desbordante possesses a slightly different set of primitives than Metanome. Currently, Desbordante supports discovery and validation for the following primitives:
\begin{enumerate}
    \item Discovery of exact functional dependencies. We support all algorithms~\cite{10.14778/2794367.2794377} that were implemented by the Metanome team, including HyFD~\cite{10.1145/2882903.2915203} and the approximate algorithm AID-FD~\cite{10.1145/2983323.2983781}.
    \item Discovery of approximate functional dependencies, using the Pyro~\cite{10.14778/3192965.3192968} and TANE~\cite{8129711} algorithms.
    \item Discovery of conditional functional dependencies using the CTANE~\cite{FanCFDDiscovering11,FanCFDDiscovering} algorithm and its variations. 
    \item Discovery of unary and n-ary Inclusion Dependencies using the Spider algortihm~\cite{Spider}.
    \item Validation of metric dependencies~\cite{4812519} (only in the console version for now).
    \item Discovery of fuzzy algebraic constraints~\cite{10.5555/1315451.1315509} (only in the console version for now).
    \item Discovery of association rules. This code was adapted from Christian Borgelt's\footnote{https://borgelt.net/fpm.html} implementations~\cite{10.1145/1133905.1133907} since it is efficient (used in the R package\footnote{https://www.rdocumentation.org/packages/arules/versions/1.6-8}) and time-proven. Following his recommendations, we have selected only ECLAT, FP-Growth and Apriori  algorithms. 
    
    \item Validation of Graph Dependencies~\cite{10.1145/2882903.2915232} (only console version for now)
    \item Naive profiling. In order to expand the userbase, we have also implemented naive profiling, which includes a number of simple statistics like min, max, number of missing values over column and so on.
\end{enumerate}

For now, we lack a significant number of primitives that Me\-ta\-no\-me has, such as UCCs, order dependencies and others. However, we have other types that are absent there (e.g. metric FDs), and which are more relevant for our use-cases. Also, to the best of our knowledge, for some of primitive types, e.g. graph dependencies, metric functional dependencies, and algebraic constraints our implementation is the only one publicly available. Finally, in the future we plan to greatly expand their number and catch up~--- some of the missing ones are already in the works.

\section{Architecture}\label{sec:architecture}

In this section, we describe the architecture of the web application that we built around its core~--- Desbordante.
Initially, Desbordante was a simple console application which used command line parameters and standard output as a user interface. In summer 2021 we have come to the vision described in the Introduction section and decided to provide it with a web interface. 

Thinking about its implementation, we have formulated the following requirements for the web application:
\begin{enumerate}
    \item Functionality. The system should be able to perform several user tasks in parallel.

    \item Recoverability. The system should be able to recover itself in case of various unexpected errors. 

    \item Efficiency and manageability. The system should be able to limit computational resources which are given to a particular user or even a task. This requirement is crucial to prevent resource overuse and will also will allow resource scheduling.

    \item Scalability. The architecture of the system should allow using several computing nodes for performing user tasks.

\end{enumerate}

Therefore, we have decided to use the microservice architecture, where each individual service performs a specific set of tasks. We have separate services for serving user requests, managing containers, executing tasks, database and task queue. Also, the microservice architecture goes well with containerization. The overall architecture showing connections between microservices is presented in Figure~\ref{fig:arch}. 

This approach also simplified dependency management as each of the services has every dependency pre-installed in its container. Thus, new versions of the application can be quickly redeployed to deliver new features to the users as soon as possible.

As the result, we have built a fault-tolerant application, which means that in case of one of the microservices failing, the application will continue its work and quickly restart the failed microservice. Our architecture is highly scalable, it can launch more task-executors if necessary.

Let us consider the architecture in detail.
\begin{enumerate}
    
    \item Serving webpages, frontend server directly interacts with the users. It is responsible for server-side rendering, a technique that moves calculations from the client browsers to the server. That speeds up page loading and makes user interface more responsible.

    \item Node.js web server provides an API that allows to run tasks, send files and receive status updates. For each task on its arrival it creates entries both in the database and the queue. Web client periodically sends pings to the server, which replies with the progress info on the task.
    After task execution is finished, it sends back either the results or an error message (if a calculation error occurred or it required too many computational resources to complete).

    \item In addition to Docker, we have our own container orchestrator that is used for executor containers management. 
    The orchestrator also behaves as a consumer for the queue. 
    For each task, it creates a new executor container and provides it with the task's payload.
    This service allows us to limit computational resources used by the executors. Additionally, in case of the executor failure it puts an error message into the database. 
    
    \item We use Kafka as a task queue. 

    That makes load balancing between multiple instances of the orchestrator service possible.

    \item PostgreSQL serves as a DBMS in our project. 
    It is used as a storage for the information about currently running or recently finished tasks, such as progress info, error messages and calculation results. Additionally it contains user profiles, session info and file metadata

    \item The kernel of the app is its task executor utilizing Desbordante as a library. Its purpose is to run the specified algorithm on the provided data. 
    On launch it acquires the data from the database and then starts the execution. While calculations are in progress, executor updates its task's status in the database. Finishing successfully or failing it returns calculation results or an error, respectively.
    Several executor containers can run at the same time, therefore making it possible to serve the users simultaneously.
    
\end{enumerate}

The main set of the microservices is accompanied by the monitoring system.
It utilizes two specific tools: Prometheus and Grafana. 
Prometheus accumulates metrics from the services, periodically collecting data from the specified endpoints. 
Grafana presents these data in the form of informative dashboards. It is also capable of notifying of events meeting the predefined requirements. 

Collected data provides some insights about the system's health, resource consumption, and execution errors. It is crucial for addressing probable hardware and software issues.

\section{Future Plans}\label{sec:future}

Desbordante is currently being actively developed. There are two primary directions: improving user experience (in a broader sense) and adding new primitives. The first one includes the following tasks:
\begin{itemize}
    \item alternative approaches to data uploading: via external file link, database connectors or import of serialized data structures (e.g. pandas pickled dataframes or NumPy arrays);
    \item export of results in most common data exchange formats;
    \item web API tokens, so remote high-performing server can be used for discovery tasks;
    \item data manipulation such as in-place table edits, column renaming or creation of new columns based on a user defined formula;
    \item regexp search over dataset cells;
    \item extending user and admin dashboard.
\end{itemize}

The second one concerns extending the pool of available primitives. There are two directions: first we plan to catch up with Metanome by adding missing primitives and at the same time we plan to continue to implement our vision and bring less known primitives to light. Near-term plans include implementing the following primitives:
\begin{itemize}
    \item matching dependencies
    \item order dependencies
    \item denial constraints
    \item differential dependencies
    \item unique column combination
    \item various types of relaxed functional dependencies
    \item graph dependency discovery 
    \item advanced dataset statistics
\end{itemize}

Finally, we also plan to touch upon system aspects:
\begin{itemize}
    \item Devise smart result caching and checking result containment.
    \item Implement stream processing and dynamic recalculation of primitives.
\end{itemize}

\section*{Acknowledgments}

We would like to thank Nikita Talalay and Bulat Biktagirov for their contribution to the project. We would also like to thank Anna Smirnova for her help with the preparation of this paper.

\section{Conclusion}

In this paper we have presented Desbordante~--- an open-source data profiler with the focus on discovery of science-intensive patterns in data. Desbordante aims to open industrial-grade primitive discovery to a broader public, focusing on domain experts who are not
IT professionals. Unlike similar systems, it is built with emphasis on industrial
application in a multi-user environment. It is efficient, resilient to crashes, and scalable. Its efficiency is ensured by implementing discovery algorithms in C++, resilience is achieved by extensive
use of containerization, and scalability is based on replication of containers.

\bibliographystyle{ACM-Reference-Format}
\bibliography{bibliography}


\begin{thebibliography}{37}


\ifx \showCODEN    \undefined \def \showCODEN     #1{\unskip}     \fi
\ifx \showDOI      \undefined \def \showDOI       #1{#1}\fi
\ifx \showISBNx    \undefined \def \showISBNx     #1{\unskip}     \fi
\ifx \showISBNxiii \undefined \def \showISBNxiii  #1{\unskip}     \fi
\ifx \showISSN     \undefined \def \showISSN      #1{\unskip}     \fi
\ifx \showLCCN     \undefined \def \showLCCN      #1{\unskip}     \fi
\ifx \shownote     \undefined \def \shownote      #1{#1}          \fi
\ifx \showarticletitle \undefined \def \showarticletitle #1{#1}   \fi
\ifx \showURL      \undefined \def \showURL       {\relax}        \fi
\providecommand\bibfield[2]{#2}
\providecommand\bibinfo[2]{#2}
\providecommand\natexlab[1]{#1}
\providecommand\showeprint[2][]{arXiv:#2}

\bibitem[\protect\citeauthoryear{Abe~Gong}{Abe~Gong}{[n.d.]}]%
        {greatExpectations}
\bibfield{author}{\bibinfo{person}{James~Campbell Abe~Gong}.}
  \bibinfo{year}{[n.d.]}\natexlab{}.
\newblock \bibinfo{title}{{Great Expectations}}.
\newblock
  \bibinfo{howpublished}{\url{https://github.com/great-expectations/great_expectations}}.
\newblock


\bibitem[\protect\citeauthoryear{Abedjan, Golab, and Naumann}{Abedjan
  et~al\mbox{.}}{2015}]%
        {10.1007/s00778-015-0389-y}
\bibfield{author}{\bibinfo{person}{Ziawasch Abedjan}, \bibinfo{person}{Lukasz
  Golab}, {and} \bibinfo{person}{Felix Naumann}.}
  \bibinfo{year}{2015}\natexlab{}.
\newblock \showarticletitle{Profiling Relational Data: A Survey}.
\newblock \bibinfo{journal}{\emph{The VLDB Journal}} \bibinfo{volume}{24},
  \bibinfo{number}{4} (\bibinfo{date}{aug} \bibinfo{year}{2015}),
  \bibinfo{pages}{557–581}.
\newblock
\showISSN{1066-8888}
\urldef\tempurl%
\url{https://doi.org/10.1007/s00778-015-0389-y}
\showDOI{\tempurl}


\bibitem[\protect\citeauthoryear{Abedjan, Golab, Naumann, and
  Papenbrock}{Abedjan et~al\mbox{.}}{2018}]%
        {10.5555/3312004}
\bibfield{author}{\bibinfo{person}{Ziawasch Abedjan}, \bibinfo{person}{Lukasz
  Golab}, \bibinfo{person}{Felix Naumann}, {and} \bibinfo{person}{Thorsten
  Papenbrock}.} \bibinfo{year}{2018}\natexlab{}.
\newblock \bibinfo{booktitle}{\emph{Data Profiling}}.
\newblock \bibinfo{publisher}{Morgan \& Claypool Publishers}.
\newblock
\showISBNx{1681734486}


\bibitem[\protect\citeauthoryear{Aggarwal and Han}{Aggarwal and Han}{2014}]%
        {10.5555/2677098}
\bibfield{author}{\bibinfo{person}{Charu~C. Aggarwal} {and}
  \bibinfo{person}{Jiawei Han}.} \bibinfo{year}{2014}\natexlab{}.
\newblock \bibinfo{booktitle}{\emph{Frequent Pattern Mining}}.
\newblock \bibinfo{publisher}{Springer Publishing Company, Incorporated}.
\newblock
\showISBNx{3319078208}


\bibitem[\protect\citeauthoryear{Bauckmann, Leser, Naumann, and
  Tietz}{Bauckmann et~al\mbox{.}}{2007}]%
        {Spider}
\bibfield{author}{\bibinfo{person}{Jana Bauckmann}, \bibinfo{person}{Ulf
  Leser}, \bibinfo{person}{Felix Naumann}, {and} \bibinfo{person}{Veronique
  Tietz}.} \bibinfo{year}{2007}\natexlab{}.
\newblock \showarticletitle{Efficiently Detecting Inclusion Dependencies}. In
  \bibinfo{booktitle}{\emph{2007 IEEE 23rd International Conference on Data
  Engineering}}. \bibinfo{pages}{1448--1450}.
\newblock
\urldef\tempurl%
\url{https://doi.org/10.1109/ICDE.2007.369032}
\showDOI{\tempurl}


\bibitem[\protect\citeauthoryear{Bertossi, Bravo, Franconi, and
  Lopatenko}{Bertossi et~al\mbox{.}}{2008}]%
        {denialConstraints}
\bibfield{author}{\bibinfo{person}{Leopoldo Bertossi}, \bibinfo{person}{Loreto
  Bravo}, \bibinfo{person}{Enrico Franconi}, {and} \bibinfo{person}{Andrei
  Lopatenko}.} \bibinfo{year}{2008}\natexlab{}.
\newblock \showarticletitle{The complexity and approximation of fixing
  numerical attributes in databases under integrity constraints}.
\newblock \bibinfo{journal}{\emph{Information Systems}} \bibinfo{volume}{33},
  \bibinfo{number}{4} (\bibinfo{year}{2008}), \bibinfo{pages}{407--434}.
\newblock
\showISSN{0306-4379}
\urldef\tempurl%
\url{https://doi.org/10.1016/j.is.2008.01.005}
\showDOI{\tempurl}
\newblock
\shownote{Selected Papers from the Tenth International Symposium on Database
  Programming Languages (DBPL 2005.}


\bibitem[\protect\citeauthoryear{Beskales, Ilyas, and Golab}{Beskales
  et~al\mbox{.}}{2010}]%
        {SAMP}
\bibfield{author}{\bibinfo{person}{George Beskales}, \bibinfo{person}{Ihab~F.
  Ilyas}, {and} \bibinfo{person}{Lukasz Golab}.}
  \bibinfo{year}{2010}\natexlab{}.
\newblock \showarticletitle{Sampling the Repairs of Functional Dependency
  Violations under Hard Constraints}.
\newblock \bibinfo{journal}{\emph{Proc. VLDB Endow.}} \bibinfo{volume}{3},
  \bibinfo{number}{1–2} (\bibinfo{date}{sep} \bibinfo{year}{2010}),
  \bibinfo{pages}{197–207}.
\newblock
\showISSN{2150-8097}
\urldef\tempurl%
\url{https://doi.org/10.14778/1920841.1920870}
\showDOI{\tempurl}


\bibitem[\protect\citeauthoryear{Bleifu\ss{}, B\"{u}low, Frohnhofen, Risch,
  Wiese, Kruse, Papenbrock, and Naumann}{Bleifu\ss{} et~al\mbox{.}}{2016}]%
        {10.1145/2983323.2983781}
\bibfield{author}{\bibinfo{person}{Tobias Bleifu\ss{}},
  \bibinfo{person}{Susanne B\"{u}low}, \bibinfo{person}{Johannes Frohnhofen},
  \bibinfo{person}{Julian Risch}, \bibinfo{person}{Georg Wiese},
  \bibinfo{person}{Sebastian Kruse}, \bibinfo{person}{Thorsten Papenbrock},
  {and} \bibinfo{person}{Felix Naumann}.} \bibinfo{year}{2016}\natexlab{}.
\newblock \showarticletitle{Approximate Discovery of Functional Dependencies
  for Large Datasets}. In \bibinfo{booktitle}{\emph{Proceedings of the 25th ACM
  International on Conference on Information and Knowledge Management}}
  \emph{(\bibinfo{series}{CIKM '16})}. \bibinfo{publisher}{Association for
  Computing Machinery}, \bibinfo{address}{New York, NY, USA},
  \bibinfo{pages}{1803–1812}.
\newblock
\showISBNx{9781450340731}
\urldef\tempurl%
\url{https://doi.org/10.1145/2983323.2983781}
\showDOI{\tempurl}


\bibitem[\protect\citeauthoryear{Borgelt}{Borgelt}{2005}]%
        {10.1145/1133905.1133907}
\bibfield{author}{\bibinfo{person}{Christian Borgelt}.}
  \bibinfo{year}{2005}\natexlab{}.
\newblock \showarticletitle{An Implementation of the FP-Growth Algorithm}. In
  \bibinfo{booktitle}{\emph{Proceedings of the 1st International Workshop on
  Open Source Data Mining: Frequent Pattern Mining Implementations}}
  \emph{(\bibinfo{series}{OSDM '05})}. \bibinfo{publisher}{Association for
  Computing Machinery}, \bibinfo{address}{New York, NY, USA},
  \bibinfo{pages}{1–5}.
\newblock
\showISBNx{1595932100}
\urldef\tempurl%
\url{https://doi.org/10.1145/1133905.1133907}
\showDOI{\tempurl}


\bibitem[\protect\citeauthoryear{Brown and Hass}{Brown and Hass}{2003}]%
        {10.5555/1315451.1315509}
\bibfield{author}{\bibinfo{person}{Paul~G. Brown} {and}
  \bibinfo{person}{Peter~J. Hass}.} \bibinfo{year}{2003}\natexlab{}.
\newblock \showarticletitle{BHUNT: Automatic Discovery of Fuzzy Algebraic
  Constraints in Relational Data}. In \bibinfo{booktitle}{\emph{Proceedings of
  the 29th International Conference on Very Large Data Bases - Volume 29}}
  \emph{(\bibinfo{series}{VLDB '03})}. \bibinfo{publisher}{VLDB Endowment},
  \bibinfo{pages}{668–679}.
\newblock
\showISBNx{0127224424}


\bibitem[\protect\citeauthoryear{Brugman}{Brugman}{2019}]%
        {pandasProfiling}
\bibfield{author}{\bibinfo{person}{Simon Brugman}.}
  \bibinfo{year}{2019}\natexlab{}.
\newblock \bibinfo{title}{{pandas-profiling: Exploratory Data Analysis for
  Python}}.
\newblock
  \bibinfo{howpublished}{\url{https://github.com/pandas-profiling/pandas-profiling}}.
\newblock


\bibitem[\protect\citeauthoryear{Caruccio, Deufemia, and Polese}{Caruccio
  et~al\mbox{.}}{2016}]%
        {7219433}
\bibfield{author}{\bibinfo{person}{Loredana Caruccio},
  \bibinfo{person}{Vincenzo Deufemia}, {and} \bibinfo{person}{Giuseppe
  Polese}.} \bibinfo{year}{2016}\natexlab{}.
\newblock \showarticletitle{Relaxed Functional Dependencies—A Survey of
  Approaches}.
\newblock \bibinfo{journal}{\emph{IEEE Transactions on Knowledge and Data
  Engineering}} \bibinfo{volume}{28}, \bibinfo{number}{1}
  (\bibinfo{year}{2016}), \bibinfo{pages}{147--165}.
\newblock
\urldef\tempurl%
\url{https://doi.org/10.1109/TKDE.2015.2472010}
\showDOI{\tempurl}


\bibitem[\protect\citeauthoryear{Chu, Ilyas, and Papotti}{Chu
  et~al\mbox{.}}{2013}]%
        {Holistic}
\bibfield{author}{\bibinfo{person}{Xu Chu}, \bibinfo{person}{Ihab~F. Ilyas},
  {and} \bibinfo{person}{Paolo Papotti}.} \bibinfo{year}{2013}\natexlab{}.
\newblock \showarticletitle{Holistic data cleaning: Putting violations into
  context}. In \bibinfo{booktitle}{\emph{2013 IEEE 29th International
  Conference on Data Engineering (ICDE)}}. \bibinfo{pages}{458--469}.
\newblock
\urldef\tempurl%
\url{https://doi.org/10.1109/ICDE.2013.6544847}
\showDOI{\tempurl}


\bibitem[\protect\citeauthoryear{Chu, Morcos, Ilyas, Ouzzani, Papotti, Tang,
  and Ye}{Chu et~al\mbox{.}}{2015}]%
        {KATARA}
\bibfield{author}{\bibinfo{person}{Xu Chu}, \bibinfo{person}{John Morcos},
  \bibinfo{person}{Ihab~F. Ilyas}, \bibinfo{person}{Mourad Ouzzani},
  \bibinfo{person}{Paolo Papotti}, \bibinfo{person}{Nan Tang}, {and}
  \bibinfo{person}{Yin Ye}.} \bibinfo{year}{2015}\natexlab{}.
\newblock \showarticletitle{KATARA: Reliable Data Cleaning with Knowledge Bases
  and Crowdsourcing}.
\newblock \bibinfo{journal}{\emph{Proc. VLDB Endow.}} \bibinfo{volume}{8},
  \bibinfo{number}{12} (\bibinfo{date}{aug} \bibinfo{year}{2015}),
  \bibinfo{pages}{1952–1955}.
\newblock
\showISSN{2150-8097}
\urldef\tempurl%
\url{https://doi.org/10.14778/2824032.2824109}
\showDOI{\tempurl}


\bibitem[\protect\citeauthoryear{Fan, Geerts, Lakshmanan, and Xiong}{Fan
  et~al\mbox{.}}{2009}]%
        {FanCFDDiscovering}
\bibfield{author}{\bibinfo{person}{Wenfei Fan}, \bibinfo{person}{Floris
  Geerts}, \bibinfo{person}{Laks V.~S. Lakshmanan}, {and} \bibinfo{person}{Ming
  Xiong}.} \bibinfo{year}{2009}\natexlab{}.
\newblock \showarticletitle{Discovering Conditional Functional Dependencies}.
  In \bibinfo{booktitle}{\emph{Proceedings of the 25th International Conference
  on Data Engineering, ICDE 2009, March 29 2009 - April 2 2009, Shanghai,
  China}}. \bibinfo{publisher}{IEEE}, \bibinfo{pages}{1231--1234}.
\newblock
\showISBNx{978-0-7695-3545-6}
\urldef\tempurl%
\url{https://doi.org/10.1109/ICDE.2009.208}
\showDOI{\tempurl}


\bibitem[\protect\citeauthoryear{Fan, Geerts, Li, and Xiong}{Fan
  et~al\mbox{.}}{2011}]%
        {FanCFDDiscovering11}
\bibfield{author}{\bibinfo{person}{Wenfei Fan}, \bibinfo{person}{Floris
  Geerts}, \bibinfo{person}{Jianzhong Li}, {and} \bibinfo{person}{Ming Xiong}.}
  \bibinfo{year}{2011}\natexlab{}.
\newblock \showarticletitle{Discovering Conditional Functional Dependencies}.
\newblock \bibinfo{journal}{\emph{IEEE Trans. Knowl. Data Eng.}}
  \bibinfo{volume}{23}, \bibinfo{number}{5} (\bibinfo{year}{2011}),
  \bibinfo{pages}{683--698}.
\newblock
\urldef\tempurl%
\url{https://doi.org/10.1109/TKDE.2010.154}
\showDOI{\tempurl}


\bibitem[\protect\citeauthoryear{Fan, Wu, and Xu}{Fan et~al\mbox{.}}{2016}]%
        {10.1145/2882903.2915232}
\bibfield{author}{\bibinfo{person}{Wenfei Fan}, \bibinfo{person}{Yinghui Wu},
  {and} \bibinfo{person}{Jingbo Xu}.} \bibinfo{year}{2016}\natexlab{}.
\newblock \showarticletitle{Functional Dependencies for Graphs}. In
  \bibinfo{booktitle}{\emph{Proceedings of the 2016 International Conference on
  Management of Data}} \emph{(\bibinfo{series}{SIGMOD '16})}.
  \bibinfo{publisher}{Association for Computing Machinery},
  \bibinfo{address}{New York, NY, USA}, \bibinfo{pages}{1843–1857}.
\newblock
\showISBNx{9781450335317}
\urldef\tempurl%
\url{https://doi.org/10.1145/2882903.2915232}
\showDOI{\tempurl}


\bibitem[\protect\citeauthoryear{{Huhtala}, {Kärkkäinen}, {Porkka}, and
  {Toivonen}}{{Huhtala} et~al\mbox{.}}{1999}]%
        {8129711}
\bibfield{author}{\bibinfo{person}{Y. {Huhtala}}, \bibinfo{person}{J.
  {Kärkkäinen}}, \bibinfo{person}{P. {Porkka}}, {and} \bibinfo{person}{H.
  {Toivonen}}.} \bibinfo{year}{1999}\natexlab{}.
\newblock \showarticletitle{Tane: An Efficient Algorithm for Discovering
  Functional and Approximate Dependencies}.
\newblock \bibinfo{journal}{\emph{Comput. J.}} \bibinfo{volume}{42},
  \bibinfo{number}{2} (\bibinfo{year}{1999}), \bibinfo{pages}{100--111}.
\newblock
\urldef\tempurl%
\url{https://doi.org/10.1093/comjnl/42.2.100}
\showDOI{\tempurl}


\bibitem[\protect\citeauthoryear{Hulsebos, Hu, Bakker, Zgraggen, Satyanarayan,
  Kraska, Demiralp, and Hidalgo}{Hulsebos et~al\mbox{.}}{2019}]%
        {10.1145/3292500.3330993}
\bibfield{author}{\bibinfo{person}{Madelon Hulsebos}, \bibinfo{person}{Kevin
  Hu}, \bibinfo{person}{Michiel Bakker}, \bibinfo{person}{Emanuel Zgraggen},
  \bibinfo{person}{Arvind Satyanarayan}, \bibinfo{person}{Tim Kraska},
  \bibinfo{person}{\c{C}agatay Demiralp}, {and} \bibinfo{person}{C\'{e}sar
  Hidalgo}.} \bibinfo{year}{2019}\natexlab{}.
\newblock \showarticletitle{Sherlock: A Deep Learning Approach to Semantic Data
  Type Detection}. In \bibinfo{booktitle}{\emph{Proceedings of the 25th ACM
  SIGKDD International Conference on Knowledge Discovery \& Data Mining}}
  \emph{(\bibinfo{series}{KDD '19})}. \bibinfo{publisher}{Association for
  Computing Machinery}, \bibinfo{address}{New York, NY, USA},
  \bibinfo{pages}{1500–1508}.
\newblock
\showISBNx{9781450362016}
\urldef\tempurl%
\url{https://doi.org/10.1145/3292500.3330993}
\showDOI{\tempurl}


\bibitem[\protect\citeauthoryear{Ilyas and Chu}{Ilyas and Chu}{2019}]%
        {10.1145/3310205}
\bibfield{author}{\bibinfo{person}{Ihab~F. Ilyas} {and} \bibinfo{person}{Xu
  Chu}.} \bibinfo{year}{2019}\natexlab{}.
\newblock \bibinfo{booktitle}{\emph{Data Cleaning}}.
\newblock \bibinfo{publisher}{Association for Computing Machinery},
  \bibinfo{address}{New York, NY, USA}.
\newblock
\showISBNx{9781450371520}


\bibitem[\protect\citeauthoryear{Jakob, Rhinelander, and Moldovan}{Jakob
  et~al\mbox{.}}{2017}]%
        {pybind11}
\bibfield{author}{\bibinfo{person}{Wenzel Jakob}, \bibinfo{person}{Jason
  Rhinelander}, {and} \bibinfo{person}{Dean Moldovan}.}
  \bibinfo{year}{2017}\natexlab{}.
\newblock \bibinfo{title}{pybind11 -- Seamless operability between C++11 and
  Python}.
\newblock
\newblock
\newblock
\shownote{https://github.com/pybind/pybind11.}


\bibitem[\protect\citeauthoryear{Koudas, Saha, Srivastava, and
  Venkatasubramanian}{Koudas et~al\mbox{.}}{2009}]%
        {4812519}
\bibfield{author}{\bibinfo{person}{Nick Koudas}, \bibinfo{person}{Avishek
  Saha}, \bibinfo{person}{Divesh Srivastava}, {and} \bibinfo{person}{Suresh
  Venkatasubramanian}.} \bibinfo{year}{2009}\natexlab{}.
\newblock \showarticletitle{Metric Functional Dependencies}. In
  \bibinfo{booktitle}{\emph{2009 IEEE 25th International Conference on Data
  Engineering}}. \bibinfo{pages}{1275--1278}.
\newblock
\urldef\tempurl%
\url{https://doi.org/10.1109/ICDE.2009.219}
\showDOI{\tempurl}


\bibitem[\protect\citeauthoryear{Kruse and Naumann}{Kruse and Naumann}{2018}]%
        {10.14778/3192965.3192968}
\bibfield{author}{\bibinfo{person}{Sebastian Kruse} {and}
  \bibinfo{person}{Felix Naumann}.} \bibinfo{year}{2018}\natexlab{}.
\newblock \showarticletitle{Efficient Discovery of Approximate Dependencies}.
\newblock \bibinfo{journal}{\emph{Proc. VLDB Endow.}} \bibinfo{volume}{11},
  \bibinfo{number}{7} (\bibinfo{date}{March} \bibinfo{year}{2018}),
  \bibinfo{pages}{759–772}.
\newblock
\showISSN{2150-8097}
\urldef\tempurl%
\url{https://doi.org/10.14778/3192965.3192968}
\showDOI{\tempurl}


\bibitem[\protect\citeauthoryear{M\"{u}ller, Castelo, Qazi, and
  Freire}{M\"{u}ller et~al\mbox{.}}{2021}]%
        {10.14778/3476311.3476339}
\bibfield{author}{\bibinfo{person}{Heiko M\"{u}ller}, \bibinfo{person}{Sonia
  Castelo}, \bibinfo{person}{Munaf Qazi}, {and} \bibinfo{person}{Juliana
  Freire}.} \bibinfo{year}{2021}\natexlab{}.
\newblock \showarticletitle{From Papers to Practice: The Openclean Open-Source
  Data Cleaning Library}.
\newblock \bibinfo{journal}{\emph{Proc. VLDB Endow.}} \bibinfo{volume}{14},
  \bibinfo{number}{12} (\bibinfo{date}{oct} \bibinfo{year}{2021}),
  \bibinfo{pages}{2763–2766}.
\newblock
\showISSN{2150-8097}
\urldef\tempurl%
\url{https://doi.org/10.14778/3476311.3476339}
\showDOI{\tempurl}


\bibitem[\protect\citeauthoryear{{``No Bugs'' Hare}}{{``No Bugs''
  Hare}}{2018}]%
        {MemoryAllocators}
\bibfield{author}{\bibinfo{person}{{``No Bugs'' Hare}}.}
  \bibinfo{year}{2018}\natexlab{}.
\newblock \bibinfo{title}{{Testing Memory Allocators: ptmalloc2 vs tcmalloc vs
  hoard vs jemalloc While Trying to Simulate Real-World Loads}}.
\newblock
\newblock
\urldef\tempurl%
\url{http://ithare.com/testing-memory-allocators-ptmalloc2-tcmalloc-hoard-jemalloc-while-trying-to-simulate-real-world-loads/}
\showURL{%
\tempurl}


\bibitem[\protect\citeauthoryear{Papenbrock, Bergmann, Finke, Zwiener, and
  Naumann}{Papenbrock et~al\mbox{.}}{2015a}]%
        {10.14778/2824032.2824086}
\bibfield{author}{\bibinfo{person}{Thorsten Papenbrock}, \bibinfo{person}{Tanja
  Bergmann}, \bibinfo{person}{Moritz Finke}, \bibinfo{person}{Jakob Zwiener},
  {and} \bibinfo{person}{Felix Naumann}.} \bibinfo{year}{2015}\natexlab{a}.
\newblock \showarticletitle{Data Profiling with {M}etanome}.
\newblock \bibinfo{journal}{\emph{Proc. VLDB Endow.}} \bibinfo{volume}{8},
  \bibinfo{number}{12} (\bibinfo{date}{Aug.} \bibinfo{year}{2015}),
  \bibinfo{pages}{1860–1863}.
\newblock
\showISSN{2150-8097}
\urldef\tempurl%
\url{https://doi.org/10.14778/2824032.2824086}
\showDOI{\tempurl}


\bibitem[\protect\citeauthoryear{Papenbrock, Ehrlich, Marten, Neubert, Rudolph,
  Sch\"{o}nberg, Zwiener, and Naumann}{Papenbrock et~al\mbox{.}}{2015b}]%
        {10.14778/2794367.2794377}
\bibfield{author}{\bibinfo{person}{Thorsten Papenbrock}, \bibinfo{person}{Jens
  Ehrlich}, \bibinfo{person}{Jannik Marten}, \bibinfo{person}{Tommy Neubert},
  \bibinfo{person}{Jan-Peer Rudolph}, \bibinfo{person}{Martin Sch\"{o}nberg},
  \bibinfo{person}{Jakob Zwiener}, {and} \bibinfo{person}{Felix Naumann}.}
  \bibinfo{year}{2015}\natexlab{b}.
\newblock \showarticletitle{Functional Dependency Discovery: An Experimental
  Evaluation of Seven Algorithms}.
\newblock \bibinfo{journal}{\emph{Proc. VLDB Endow.}} \bibinfo{volume}{8},
  \bibinfo{number}{10} (\bibinfo{date}{June} \bibinfo{year}{2015}),
  \bibinfo{pages}{1082–1093}.
\newblock
\showISSN{2150-8097}
\urldef\tempurl%
\url{https://doi.org/10.14778/2794367.2794377}
\showDOI{\tempurl}


\bibitem[\protect\citeauthoryear{Papenbrock and Naumann}{Papenbrock and
  Naumann}{2016}]%
        {10.1145/2882903.2915203}
\bibfield{author}{\bibinfo{person}{Thorsten Papenbrock} {and}
  \bibinfo{person}{Felix Naumann}.} \bibinfo{year}{2016}\natexlab{}.
\newblock \showarticletitle{A Hybrid Approach to Functional Dependency
  Discovery}. In \bibinfo{booktitle}{\emph{Proceedings of the 2016
  International Conference on Management of Data}}
  \emph{(\bibinfo{series}{SIGMOD '16})}. \bibinfo{publisher}{Association for
  Computing Machinery}, \bibinfo{address}{New York, NY, USA},
  \bibinfo{pages}{821–833}.
\newblock
\showISBNx{9781450335317}
\urldef\tempurl%
\url{https://doi.org/10.1145/2882903.2915203}
\showDOI{\tempurl}


\bibitem[\protect\citeauthoryear{Rekatsinas, Chu, Ilyas, and R\'{e}}{Rekatsinas
  et~al\mbox{.}}{2017}]%
        {HoloClean}
\bibfield{author}{\bibinfo{person}{Theodoros Rekatsinas}, \bibinfo{person}{Xu
  Chu}, \bibinfo{person}{Ihab~F. Ilyas}, {and} \bibinfo{person}{Christopher
  R\'{e}}.} \bibinfo{year}{2017}\natexlab{}.
\newblock \showarticletitle{HoloClean: Holistic Data Repairs with Probabilistic
  Inference}.
\newblock \bibinfo{journal}{\emph{Proc. VLDB Endow.}} \bibinfo{volume}{10},
  \bibinfo{number}{11} (\bibinfo{date}{aug} \bibinfo{year}{2017}),
  \bibinfo{pages}{1190–1201}.
\newblock
\showISSN{2150-8097}
\urldef\tempurl%
\url{https://doi.org/10.14778/3137628.3137631}
\showDOI{\tempurl}


\bibitem[\protect\citeauthoryear{Rezig, Ouzzani, Aref, Elmagarmid, Mahmood, and
  Stonebraker}{Rezig et~al\mbox{.}}{2021}]%
        {Horizon}
\bibfield{author}{\bibinfo{person}{El~Kindi Rezig}, \bibinfo{person}{Mourad
  Ouzzani}, \bibinfo{person}{Walid~G. Aref}, \bibinfo{person}{Ahmed~K.
  Elmagarmid}, \bibinfo{person}{Ahmed~R. Mahmood}, {and}
  \bibinfo{person}{Michael Stonebraker}.} \bibinfo{year}{2021}\natexlab{}.
\newblock \showarticletitle{Horizon: Scalable Dependency-Driven Data Cleaning}.
\newblock \bibinfo{journal}{\emph{Proc. VLDB Endow.}} \bibinfo{volume}{14},
  \bibinfo{number}{11} (\bibinfo{date}{oct} \bibinfo{year}{2021}),
  \bibinfo{pages}{2546–2554}.
\newblock
\showISSN{2150-8097}
\urldef\tempurl%
\url{https://doi.org/10.14778/3476249.3476301}
\showDOI{\tempurl}


\bibitem[\protect\citeauthoryear{Saxena, Golab, and Ilyas}{Saxena
  et~al\mbox{.}}{2019}]%
        {10.14778/3342263.3342638}
\bibfield{author}{\bibinfo{person}{Hemant Saxena}, \bibinfo{person}{Lukasz
  Golab}, {and} \bibinfo{person}{Ihab~F. Ilyas}.}
  \bibinfo{year}{2019}\natexlab{}.
\newblock \showarticletitle{Distributed Implementations of Dependency Discovery
  Algorithms}.
\newblock \bibinfo{journal}{\emph{Proc. VLDB Endow.}} \bibinfo{volume}{12},
  \bibinfo{number}{11} (\bibinfo{date}{jul} \bibinfo{year}{2019}),
  \bibinfo{pages}{1624–1636}.
\newblock
\showISSN{2150-8097}
\urldef\tempurl%
\url{https://doi.org/10.14778/3342263.3342638}
\showDOI{\tempurl}


\bibitem[\protect\citeauthoryear{Song and He}{Song and He}{2021}]%
        {10.1145/3448016.3457250}
\bibfield{author}{\bibinfo{person}{Jie Song} {and} \bibinfo{person}{Yeye He}.}
  \bibinfo{year}{2021}\natexlab{}.
\newblock \showarticletitle{Auto-Validate: Unsupervised Data Validation Using
  Data-Domain Patterns Inferred from Data Lakes}. In
  \bibinfo{booktitle}{\emph{Proceedings of the 2021 International Conference on
  Management of Data}} \emph{(\bibinfo{series}{SIGMOD '21})}.
  \bibinfo{publisher}{Association for Computing Machinery},
  \bibinfo{address}{New York, NY, USA}, \bibinfo{pages}{1678–1691}.
\newblock
\showISBNx{9781450383431}
\urldef\tempurl%
\url{https://doi.org/10.1145/3448016.3457250}
\showDOI{\tempurl}


\bibitem[\protect\citeauthoryear{Song, Gao, Huang, and Wang}{Song
  et~al\mbox{.}}{2022}]%
        {9302878}
\bibfield{author}{\bibinfo{person}{Shaoxu Song}, \bibinfo{person}{Fei Gao},
  \bibinfo{person}{Ruihong Huang}, {and} \bibinfo{person}{Chaokun Wang}.}
  \bibinfo{year}{2022}\natexlab{}.
\newblock \showarticletitle{Data Dependencies Extended for Variety and
  Veracity: A Family Tree}.
\newblock \bibinfo{journal}{\emph{IEEE Transactions on Knowledge and Data
  Engineering}} \bibinfo{volume}{34}, \bibinfo{number}{10}
  (\bibinfo{year}{2022}), \bibinfo{pages}{4717--4736}.
\newblock
\urldef\tempurl%
\url{https://doi.org/10.1109/TKDE.2020.3046443}
\showDOI{\tempurl}


\bibitem[\protect\citeauthoryear{Strutovskiy, Bobrov, Smirnov, and
  Chernishev}{Strutovskiy et~al\mbox{.}}{2021}]%
        {9435469}
\bibfield{author}{\bibinfo{person}{Maxim Strutovskiy}, \bibinfo{person}{Nikita
  Bobrov}, \bibinfo{person}{Kirill Smirnov}, {and} \bibinfo{person}{George
  Chernishev}.} \bibinfo{year}{2021}\natexlab{}.
\newblock \showarticletitle{Desbordante: a Framework for Exploring Limits of
  Dependency Discovery Algorithms}. In \bibinfo{booktitle}{\emph{2021 29th
  Conference of Open Innovations Association (FRUCT)}}.
  \bibinfo{pages}{344--354}.
\newblock
\urldef\tempurl%
\url{https://doi.org/10.23919/FRUCT52173.2021.9435469}
\showDOI{\tempurl}


\bibitem[\protect\citeauthoryear{Tsurinov, Shpynov, Lukashina, Likholetova, and
  Artyomov}{Tsurinov et~al\mbox{.}}{2021}]%
        {10.1145/3459930.3469499}
\bibfield{author}{\bibinfo{person}{Petr Tsurinov}, \bibinfo{person}{Oleg
  Shpynov}, \bibinfo{person}{Nina Lukashina}, \bibinfo{person}{Daria
  Likholetova}, {and} \bibinfo{person}{Maxim Artyomov}.}
  \bibinfo{year}{2021}\natexlab{}.
\newblock \showarticletitle{FARM: Hierarchical Association Rule Mining and
  Visualization Method}. In \bibinfo{booktitle}{\emph{Proceedings of the 12th
  ACM Conference on Bioinformatics, Computational Biology, and Health
  Informatics}} \emph{(\bibinfo{series}{BCB '21})}.
  \bibinfo{publisher}{Association for Computing Machinery},
  \bibinfo{address}{New York, NY, USA}, Article \bibinfo{articleno}{70},
  \bibinfo{numpages}{1}~pages.
\newblock
\showISBNx{9781450384506}
\urldef\tempurl%
\url{https://doi.org/10.1145/3459930.3469499}
\showDOI{\tempurl}


\bibitem[\protect\citeauthoryear{Yakout, Berti-\'{E}quille, and
  Elmagarmid}{Yakout et~al\mbox{.}}{2013}]%
        {SCARE}
\bibfield{author}{\bibinfo{person}{Mohamed Yakout}, \bibinfo{person}{Laure
  Berti-\'{E}quille}, {and} \bibinfo{person}{Ahmed~K. Elmagarmid}.}
  \bibinfo{year}{2013}\natexlab{}.
\newblock \showarticletitle{Don't Be SCAREd: Use SCalable Automatic REpairing
  with Maximal Likelihood and Bounded Changes}. In
  \bibinfo{booktitle}{\emph{Proceedings of the 2013 ACM SIGMOD International
  Conference on Management of Data}} \emph{(\bibinfo{series}{SIGMOD '13})}.
  \bibinfo{publisher}{Association for Computing Machinery},
  \bibinfo{address}{New York, NY, USA}, \bibinfo{pages}{553–564}.
\newblock
\showISBNx{9781450320375}
\urldef\tempurl%
\url{https://doi.org/10.1145/2463676.2463706}
\showDOI{\tempurl}


\bibitem[\protect\citeauthoryear{Yu and Sun}{Yu and Sun}{1989}]%
        {algebraicConstraints}
\bibfield{author}{\bibinfo{person}{C.T. Yu} {and} \bibinfo{person}{W. Sun}.}
  \bibinfo{year}{1989}\natexlab{}.
\newblock \showarticletitle{Automatic knowledge acquisition and maintenance for
  semantic query optimization}.
\newblock \bibinfo{journal}{\emph{IEEE Transactions on Knowledge and Data
  Engineering}} \bibinfo{volume}{1}, \bibinfo{number}{3}
  (\bibinfo{year}{1989}), \bibinfo{pages}{362--375}.
\newblock
\urldef\tempurl%
\url{https://doi.org/10.1109/69.87981}
\showDOI{\tempurl}


\end{thebibliography}

%
% ...

\end{document}